\documentclass[twocolumn,prl,aps]{revtex4}

\usepackage{epsf}
\usepackage{dcolumn}
\usepackage{epsf}

\begin{document}
\draft
\title{Sub-Poissonian Shot Noise in Molecular Wires}
\author{S. Dallakyan and S. Mazumdar}
\address{Department of Physics, University of Arizona, Tucson, AZ, 85721, USA}
\date{\today}
\begin{abstract}
We investigate the transport behavior of polyene molecules sandwiched between
two metallic contacts using the non-equilibrium Green's
function formalism. We calculate both current and noise power as a function
of applied voltage and
show that they decrease with increasing size of the polyene molecules.
We find that even with symmetric
connection to metallic contacts, current verus voltage
curves can be asymmetric for asymmetrically substituted polyenes.
Most importantly, we demonstrate a cross-over from Poissonian to
sub-Poissonian behavior in the shot noise as a function of applied voltage.
The algorithm for noise power calculation can be used for designing
molecules with low noise.
\end{abstract}
\maketitle

Recent progress in molecular electronics has led to many interesting experiments
on single-molecule junctions \cite{N1}. In particular, current-voltage (I-V) curves for a number of molecules %
have been measured in lithographically fabricated break junctions \cite{C11} and using
scanning tunneling microscopy (STM) \cite{C12}.
Several different theoretical approaches have been suggested for modeling
charge-transport across single molecules \cite{C2}, in conjugation with molecular modeling at
various different levels of sophistication,
ranging from the
simple H\"{u}ckel model
\cite{R1} to density functional theories \cite{R2}.
The nonequilibrium Green's Function (NEGF) formalism \cite{M1},
which allows self-consistent calculations of the electrostatic potential
along the junction and transmission probabilities, offers a very promising method
of calculating I-V characteristics of single molecule junctions.

Existing calculations of transport in organic molecular junctions have mostly
been limited to determinations of I-V characteristics.
It is well known, however, that thermal and quantum fluctuations play
important roles in the nanometer scale. For future device applications,
it therefore is essential to
investigate current fluctuations in these systems.
In order to design molecular junctions
with the lowest possible current fluctuations, an efficient algorithm to
calculate the noise power in molecular junctions is needed.
To the best of our knowledge, there exists curently only one such calculation of
the shot noise, in Si atomic wires, using a field-theoretic approach \cite{R22}.
In the present Letter we investigate
polyene molecules sandwiched between two metallic contacts using the NEGF
technique. We calculate both the current and the noise
power as a function of the applied voltage. The latter is calculated using an
an expression due to Blanter and B\"{u}ttiker \cite{M2}.
Our I-V curves shows the step-like behavior that has been
observed in resonant tunneling junctions \cite{M2}. We show that the current
and noise power decreases with
increasing size of the polyene molecules.
We also show that I-V characteristics can be asymmetric
even with symmetric connections to the metal, provided that the molecule under
investigation is
asymmetric \cite{C11}.
One of the most interesting results of our calculation is Poissonian to
sub-Poissonian
crossover in the shot noise as a function of the applied voltage,
which we hope can be tested experimentally.

The expressions for the current $\langle I \rangle$ and the noise power $S$ for the general
two-terminal device are given by \cite{M2}

\begin{equation}
\label{R1}
\langle I \rangle = \frac{2e}{h}\int_{-\infty}^\infty dE~T(E)~(f_L-f_R),
\end{equation}

\begin{eqnarray}
\label{R2}
S \equiv \frac{1}{2}\int_{-\infty}^\infty dt \langle \Delta I(t)\Delta I(0) + \Delta I(0)\Delta I(t) \rangle\nonumber\\
=\frac{4e^2}{h}\int_{-\infty}^\infty dE\{T(E)[f_L(1-f_L)+f_R(1-f_R)]\nonumber\\
+T(E)[1-T(E)](f_L-fR)^2\},
\end{eqnarray}
where $T(E)$ is the transmission probability,
$f_L$ and $f_R$ are the Fermi distributions for left and right
terminals, respectively, and $\Delta I(t) = I(t) - \langle I \rangle$. Note the additional factor of two due to the
summation over spin
degrees of freedom.

The principal problem in calculating transmission probabilities through
a metal-molecule-metal
system is incorporating the infinitely many degrees of freedom of the
metallic contacts. The NEGF formalism allows
tracing out these degrees of freedom, such that the dimension of the Green's
function for the metal-molecule-metal system is comparable to the
degrees of freedom of the
molecule \cite{M1}. According to the NEGF formalism the transmission probability
is given by
\begin{equation}
\label{R3}
T(E) = {\rm trace} [ \Gamma_L G \Gamma_R G^\dagger],
\end{equation}
where $G=[(E+0^+)I-H-\Sigma_L-\Sigma_R]^{-1}$ is the Green's function,
$\Gamma_{L/R} = i[\Sigma_{L/R}-\Sigma^\dagger_{L/R}]$ and $\Sigma_{L/R}$ is
the self-energy of the
left/right terminal.

The molecular systems we consider are linear polyenes, described within the
simple H\"uckel Hamiltonian,
\begin{equation}
\label{R4}
H = -\sum_{i, \sigma} \epsilon_i c_{i,\sigma}^\dagger c_{i,\sigma} -
\sum_{i,\sigma}t_{i,i+1}(c_{i,\sigma}^\dagger c_{i+1,\sigma} + h.c.) \\
\end{equation}
where $c_{i,\sigma}^\dagger$ creates an electron with spin $\sigma$
on site $i$ of a linear chain and $\epsilon_i$ is the site energy.
The nearest neighbor electron hopping integrals $t_{i,i+1}$ correspond to
the alternate double and single bonds in the polyene molecules, and
are taken to be -- 2.6 eV and -- 2.2 eV, respectively. The site energies are
equal for the simple polyene (and hence may be taken to be zero). We will simulate
asymmetry effects (see below) by considering unequal site energies, which can
be achieved experimentally by chemical substituition.
Self-energies that account
for the semi-infinite leads with outgoing plane waves are now
calculated using the procedure
described in Reference \cite{M1}.
We set the
hopping integral within %
the lead -- 5 eV, the Fermi energy to 5.5 eV and temperature 300 K.
This value of the Fermi energy
corresponds to that of gold electrodes. While we report the results of calculations
done with this one set of parameters only, we emphasize that
our calculations were done with
the lead hopping integral ranging from -3.5 eV to - 5.5 eV, Fermi energy ranging from 3.5 eV - 5.5 eV and temperature 1 K - 350 K.
In all cases our results were qualitatively the same.

Assuming that the voltage drops linearly along the polyene molecule,
which is a good approximation for molecules with small
cross section \cite{C3},
we have calculated the current and noise power for
polyenes with 4 and 10 carbon atoms (butadiene and decapentaene). In Fig. 1(a) we plot
current and noise power as a function of the bias voltage ($V_b$) for butadiene.

The step-like behavior for the current has been observed in a number of mesoscopic conductors and
is a result of resonant tunneling phenomenon \cite{M2}. Since electrons can travel from one
terminal to another using molecular levels only, a sharp increase in current occurs whenever
Fermfi level of the metal aligns with the molecular level.

Our results for the noise power  indicates Poissonian ($S = 2eI$)
to sub-Poissonian ($S < 2eI$) shot noise crossover as a function
of the applied voltage. We note that sub-Poissonian shot noise has also been observed
in experiments on mesoscopic conductors \cite{M2}.

\begin{figure}[ht]
\begin{center}
\begin{tabular}{c}
\epsfysize=7.cm\epsfbox{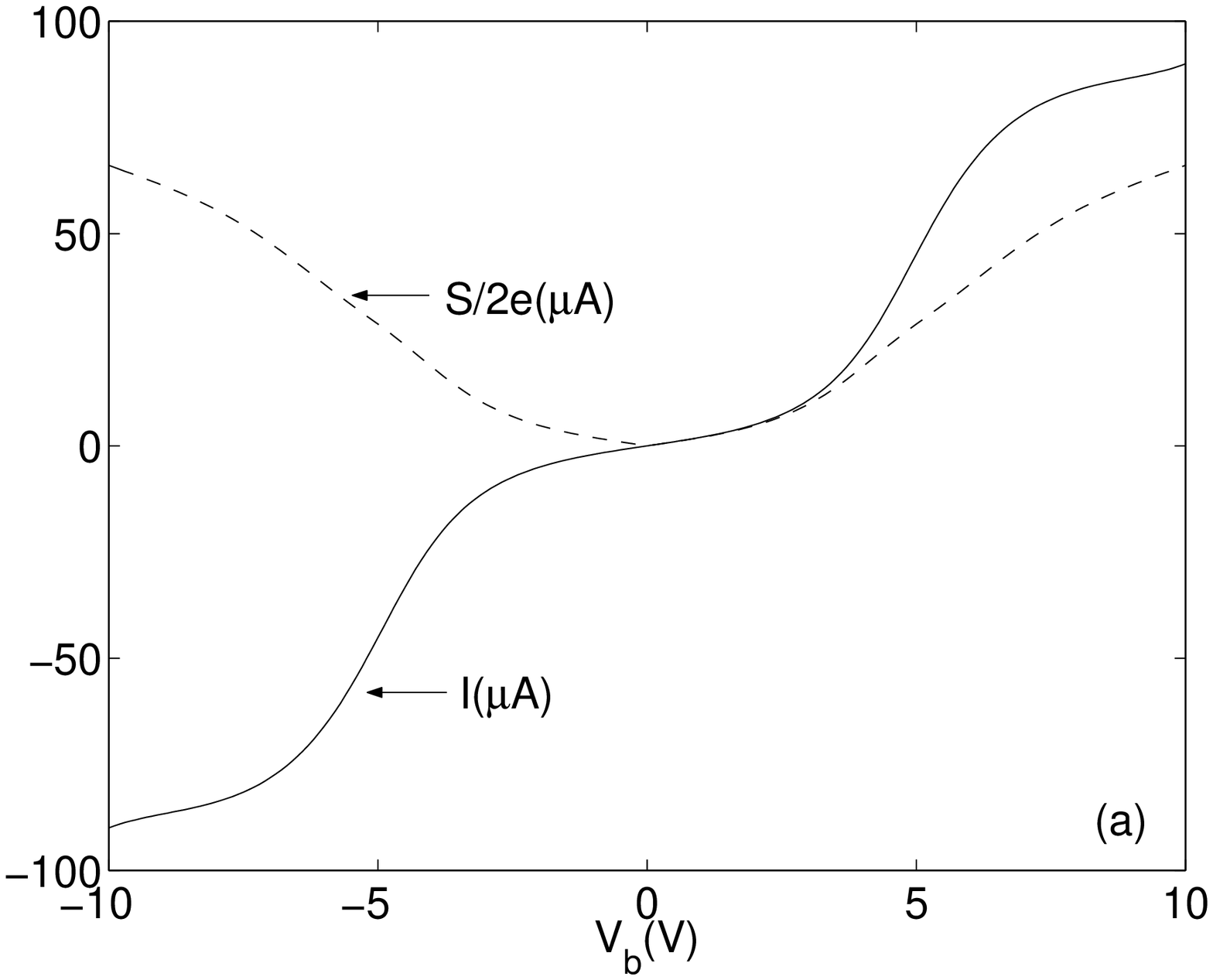} \\
\epsfysize=7cm\epsfbox{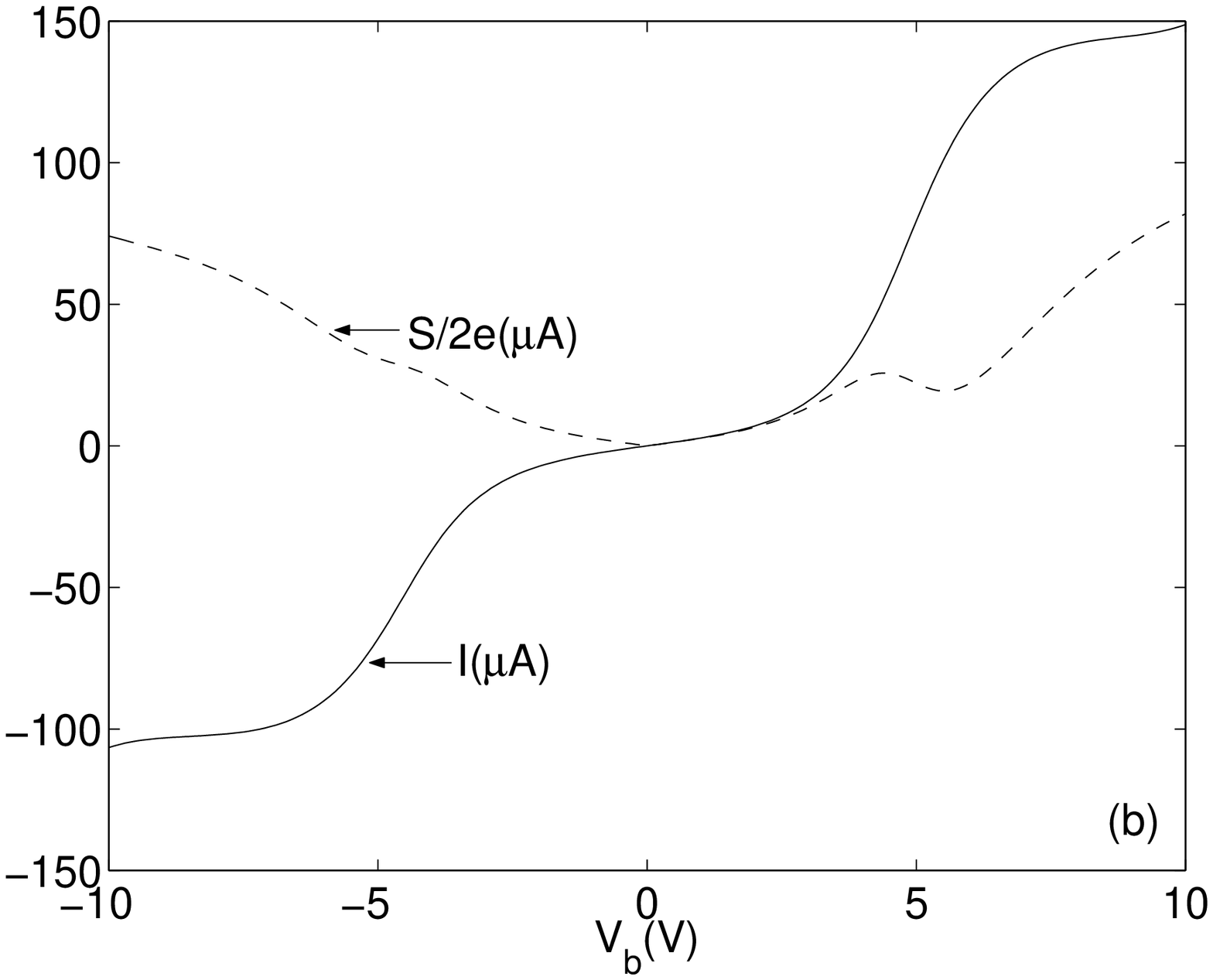} \\
\end{tabular}
\caption{Current and noise power for butadiene.
(a) symmetric molecule,
(b) asymmetric molecule.}
\label{figone}
\end{center}
\end{figure}
We introduced asymmetry in the butadiene molecule by
taking $\epsilon_1 = \epsilon_2 = -0.5$
eV and $\epsilon_3 = \epsilon_4 = 0.5$ eV.
This resulted in asymmetric I-V and noise curves (see Figs. 1(b)).
Note that for positive voltages, noise power exhibits a nonmonotonic behavior.
It would be interesting to have experimental confirmations of this result.

In Reference \cite{R3},
a different mechanism for asymmetric I-V curves,
based on asymmetric metallic contacts, was
proposed. A recent experiment has demonstrated asymmetric I-V curves for an asymmetric
molecule \cite{C11}. Thus, our caculations confirms theoretical findings in Reference
\cite{R4}, that asymmetry in I-V curves can result from asymmetry of the molecule too.

Our results for the polyene with 10 carbon atoms (decapentaene)
are shown in Fig. 2. Both the current and noise power are now smaller. In addition,
many more steps now appear in the I-V curve. This can be easily explained based on a resonant
tunneling phenomena. Since decapentaene has more energy levels that butadiene, there are more
values of the applied voltage when Fermi level of the metal aligns with the molecular levels.

In Reference \cite{M4}, where similar
I-V curves were calculated, it was suggested that low temperatures I-V
curves can be used for predicting the electronic structure of molecules, based on the number of
steps in I-V curve. Note that the plot of the noise power  indicates Poissonian
to sub-Poissonian shot noise crossover similar to one we have found for butadiene.

\begin{figure}
\begin{center}
\epsfysize=7cm\epsfbox{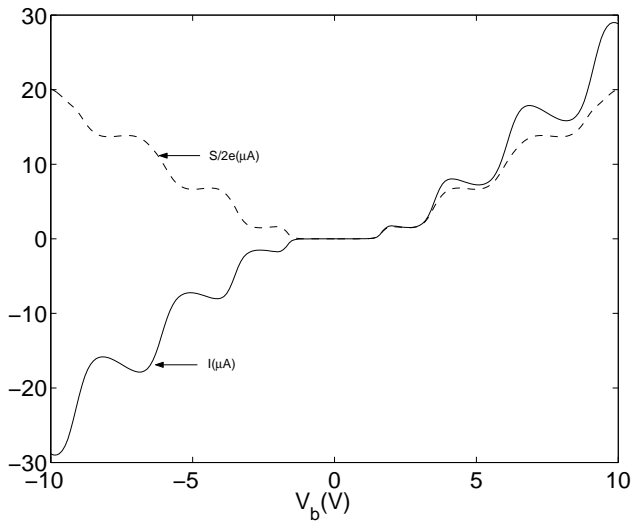}
\caption{Current and noise power for polyene with 10 carbon atoms.}
\label{figtwo}
\end{center}
\end{figure}

In summary, we have calculated current and noise power for symmetric and
asymmetric polyene molecules sandwiched between
two metallic contacts using the NEGF formalism.
Our results indicate that the current and noise power decrease
with increasing size of polyene molecules. We have shown that asymmetry
in I-V and noise power
curves can result from asymmetry of the molecule.
We have also shown Poissonian to sub-Possonian shot noise
crossover as a function
of the applied voltage. With constantly shrinking size of
electrical circuits, the limit
of single molecular junctions is being rapidly approached.
For single-molecule junctions experminetal results on the noise power are
as important as I-V characyeristics.
We believe that the algorithm presented here for the noise power calculation
can provide an efficient theoretical approach to
designing molecules with the lowest possible noise.

This work was supported by NSF DMR-0101659 and NSF ECS-0108696.

\end{document}